\documentclass[journal,twoside,web]{ieeecolor}
\usepackage{jsen}
\usepackage{cite}
\usepackage{amsmath,amssymb,amsfonts}
\usepackage{algorithm}
\usepackage{algpseudocode}
\usepackage{graphicx}
\usepackage{textcomp}
\usepackage{wrapfig}
\usepackage{caption}
\usepackage{multirow}
\usepackage{makecell}
\usepackage{hyperref}
\usepackage{subcaption}

\def\BibTeX{{\rm B\kern-.05em{\sc i\kern-.025em b}\kern-.08em
    T\kern-.1667em\lower.7ex\hbox{E}\kern-.125emX}}
\definecolor{abstractbg}{rgb}{0.89804,0.94510,0.83137}
\begin{document}
\title{EEATC: A Novel Calibration Approach for Low-cost Sensors \\ IEEE-copyrighted material refer doi 10.1109/JSEN.2023.3304366 }
\author{  M V Narayana, Devendra Jalihal, \IEEEmembership{Member, IEEE}, Shiva Nagendra S M
\thanks{ M V Narayana, is a graduate student in the Department of Electrical Engineering, Indian Institute of Technology Madras, Chennai, India-600036  (e-mail: ee18d302@smail.iitm.ac.in).}
\thanks{Devendra Jalihal, is a professor in the Department of Electrical Engineering, Indian Institute of Technology Madras, Chennai, India-600036 (e-mail: dj@iitm.ac.in).}
\thanks{Shiva Nagendra S M, is a professor in the Department of Civil Engineering, Indian Institute of Technology Madras, Chennai, India-600036 (e-mail: snagendra@iitm.ac.in).}
}
\IEEEtitleabstractindextext{%
\fcolorbox{abstractbg}{abstractbg}{%
\begin{minipage}{\textwidth}%
\begin{wrapfigure}[13]{r}{5 in}%
\includegraphics[scale=0.35]{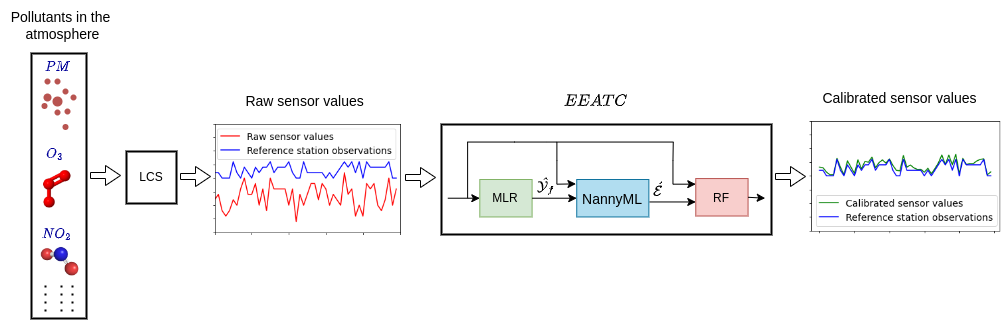}%
\end{wrapfigure}%

\begin{abstract}
Low-cost sensors (LCS) are affordable, compact, and often portable devices designed to measure various environmental parameters, including air quality. These sensors are intended to provide accessible and cost-effective solutions for monitoring pollution levels in different settings, such as indoor, outdoor and moving vehicles. However, the data produced by  LCS  is prone to various sources of error that can affect accuracy. Calibration is a well-known procedure to improve the reliability of the data produced by LCS, and several developments and efforts have been made to calibrate the LCS. This work proposes a novel Estimated Error Augmented Two-phase Calibration (\textit{EEATC}) approach to calibrate the LCS in stationary and mobile deployments. In contrast to the existing approaches, the \textit{EEATC}  calibrates the LCS in two phases, where the error estimated in the first phase calibration is augmented with the input to the second phase, which helps the second phase to learn the distributional features better to produce more accurate results. We show that the \textit{EEATC} outperforms well-known single-phase calibration models such as linear regression models (single variable linear regression (SLR) and multiple variable linear regression (MLR)) and Random forest (RF) in stationary and mobile deployments. To test the \textit{EEATC} in stationary deployments, we have used the Community Air Sensor Network (CAIRSENSE) data set approved by the United States Environmental Protection Agency (USEPA), and the mobile deployments are tested with the real-time data obtained from SensurAir, an LCS device developed and deployed on moving vehicle in Chennai, India. 
\end{abstract}

\end{minipage}}}
\maketitle

\section{Introduction}
\label{sec:introduction}
\IEEEPARstart{L}
ow-cost sensors (LCS) have gained significant attention in recent years as valuable tools for air quality monitoring (AQM). These sensors offer several advantages, including fine-granular measurements, cost-effectiveness, and portability, making them more popular than conventional reference-grade air quality monitoring instruments. These advantages enable them to use in various AQM applications.

LCS have been successfully employed in stationary deployments, where they are fixed to a specific place for an extended period. These applications include monitoring air quality in urban areas, indoor environments, or industrial sites. LCS in stationary applications can provide continuous data collection and long-term trend analysis \cite{Weissert2017, ijsis_article}. LCS are also suitable for mobile applications, where they are mounted on moving vehicles such as cars, bicycles, or drones. This mobility enables monitoring air quality or other environmental factors in different locations, offering a more comprehensive understanding of spatial variations\cite{mobilemonitoing2019}.

LCS for AQM are designed to detect and measure various pollutants present in the environment, such as particulate matter (PM), gases like nitrogen dioxide (NO2), carbon monoxide (CO), and volatile organic compounds (VOCs). They operate on different principles, including optical, electrochemical, and metal-oxide-based sensing mechanisms.

Despite these advantages, LCS data may be prone to various sources of error that can affect accuracy. Some common error sources include temperature variations, relative humidity changes, and cross-sensitivities \cite{maag_predep}. Proper calibration is crucial to address these issues and improve the accuracy of LCS data \cite{Feenstra_12lcs}. Calibration involves adjusting the sensor readings to align with known reference values. This process helps account for sensor drift, non-linearity, and other factors that can introduce inaccuracies. By calibrating LCS, the impact of error sources can be minimized, leading to more reliable and precise measurements.

Calibration approaches for low-cost sensors (LCS) have been an active area of research, and several developments and efforts have been made to enhance the accuracy of LCS through calibration, and the notable advancements in LCS calibration are as follows:

\subsubsection*{Laboratory and field studies} Extensive laboratory and field studies have been conducted to collect data from LCS in various environments \cite{Li_labstudy, Ahn_labstudy, Kuula_labstudy, Lewis_labstudy, Badura_field, Austin_labstudy}. This collected data allows for the characterization of sensor responses under varying conditions and serve as valuable inputs for developing calibration approaches. At the same, by comparing these LCS measurements with reference instruments or established air quality monitoring stations, the performance of the LCS can be validated.
\subsubsection*{Correction factor approaches} 
The collected data in the field and laboratory makes it possible to develop correction factors for sensor response, which helps to correct the sensor response biases. Temperature baseline correction and humidity baseline correction are the popular correction factor approaches \cite{Hauck2004_correction, Sun_correction, Wei_correction, samad_correction, Zheng_correction, Martin_correction, Masson_correction, Malings_correction, Popoola_correction, Rogulski_correction, Kosmopoulos2020_correction, Budde_correction, Barkjohn_temp_rh_correction, Crilley_rhcorrection_opc}. In contrast to the static correction factors approach, Aurora \cite{Aurora_correction} first modelled the sensor response under various temperature and relative humidity values and developed a self-correction algorithm in order to calibrate the metal oxide LCS.
\subsubsection*{Data driven approaches}
In the data-driven approach, LCS are calibrated with the data obtained from real-time deployments against the reference station observations, often including other influencing parameters such as temperature and humidity. This approach is more prevalent due to its simplicity, and the proliferation of machine learning (ML) and artificial intelligence (AI) catalyzed its adoption further. 
Various ML and AI models are explored to calibrate LCS. These models include single-variable linear regression (SLR) \cite{Spinelle_calibration, Hasenfratz_lr, Saukh_lr, lin_lr}, multiple-variable linear regression (MLR) \cite{KUMAR_sl_ml_knn, VanZoest_mlr}, K-nearest neighbours (K-NN) \cite{KUMAR_sl_ml_knn}, random forest (RF) \cite{Zimmerman2018, wang_rf, Han_sl_ml_rf_lstm}, artificial neural networks (ANN) \cite{Badura_sl_ml_ann, Han_sl_ml_rf_lstm, Cheng_ann, Alhasa_ann, Spinelle_lr_mlr_ann}, support vector machine (SVM) \cite{Bigi_ml_svr_rf, Mahajan_svm} extreme gradient boosting (XGB) \cite{Si_xgboost}, generalised additive model (GAM) \cite{Munir_lr_gam}. 
\subsubsection*{Hybrid models} In addition, a few works explored hybrid models obtained by combining two or more architectures to calibrate the LCS. Hagan et al. \cite{Hagan_hybrid} and Malings et al. \cite{Malings_hybrid} combined the regression model with the k-nearest neighbours and random forest, respectively, to improve the calibration accuracy compared to the stand-alone models. A similar approach that combines regression models with ANN was implemented by  Cordero et al. \cite{Cordero_hybrid}. Instead of combining the models, Ferrer-Cid et al. used multi-sensor data fusion techniques to calibrate the LCS \cite{Ferrer-Cid_data_fusion}.   

However, the calibration approaches discussed above are developed for stationary applications, which cannot be adopted directly to mobile applications due to additional error sources, such as mobility effects and rapid variation of the inputs \cite{Genikomsakis2018}. Dong et al. \cite{Dong_mobile_ann} identified that vehicle starts and stops, creates  variations in the airflow at the sensors and causes irregular measurements. They used GPS loggers to identify the samples of the start and stop positions and then filtered out the respective samples before calibrating the LCS using SVM and ANN models. Lin et al. \cite{Lin2018} applied a moving average filter to remove the noise pertained due to the vehicle's movement. In a recent study, Arman et al. \cite{Arman} highlight the application of wavelet transform to address the noise components induced in the LCS data during mobile air quality monitoring. By utilizing wavelet transform, they aimed to remove higher frequency components, which are often attributed to the acceleration and deceleration of the vehicle carrying the LCS. The denoised data, which represents the lower frequency components associated with the air quality measurements, was then used to develop land-use regression models.

Furthermore, adaptive calibration algorithms have indeed been developed to calibrate the LCS in mobile applications. These algorithms utilize feedback from sensor readings, environmental conditions, and historical data to continuously refine the calibration process in real time. For instance, Esposito et al., \cite{Esposito_dynamic_nn} proposed dynamic neural network architectures which leverage the past samples of both sensors and reference instruments in order to calibrate the LCS. Another study by Hasenfratz et al. \cite{HASENFRATZ2015268} focused on preparing urban air pollution maps using mobile sensor nodes. They employed land-use regression models but found that the accuracy of these models decreased over a longer duration of deployment. To address this issue, they incorporated calibration using past samples, which helped improve the accuracy of the models by accounting for temporal variations and sensor drift.

In this work, we propose a novel Estimated Error Augmented Two-phase Calibration (\textit{EEATC}) approach, which combines the advantages of hybrid and adaptive data-driven calibration approaches to calibrate the LCS in both stationary and mobile deployments. To the best of our knowledge, \textit{EEATC} is the first-of-its-kind calibration approach that works for both stationary as well as mobile deployments.

\textit{EEATC} calibrates the LCS in two phases, where the LCS are first calibrated with a machine learning model, which is then used to generate the calibrated output. The absolute errors obtained between the calibrated output of the first phase and reference observations are fed to nannyML, an error estimation model. The nannyML learns the error distribution and produces estimated errors. The estimated errors and the LCS data are given to the second phase, where another machine-learning model is trained to calibrate the LCS in the final stage. Since the error (the uncertainty of the calibrated values) is learned in the second phase, which helps to produce better calibration results in the second phase. 

We test the \textit{EEATC} on LCS data in stationary deployments using the Community Air Sensor Network (CAIRSENSE) data set approved by the United States Environmental Protection Agency (USEPA). In order to test \textit{EEATC} on mobile LCS data, we conducted a mobile monitoring pilot by deploying SensurAir, an LCS device that we developed, on a moving vehicle which opportunistically moves in parts of Chennai, India. 

Contributions of this work are summarized as follows,

\begin{itemize}
    \item It proposes \textit{EEATC} a novel calibration approach which calibrates the LCS in both stationary and mobile deployments. 
    \item exhaustively Tests the \textit{EEATC} on both stationary and mobile deployment data sets. 
   \item It describes a new design and deployment of an LCS device, SensurAir,  for mobile AQM monitoring. 
   \item It contributes new data sets for mobile AQM with LCS to the research community 
    \item Compares the performance of \textit{EEATC} with the state-of-art calibration models such as MLR and RF
\end{itemize}

Details of the rest of the paper are as follows. Sec. \ref{sec:Calibration_process} illustrates the \textit{EEATC}, the proposed calibration approach. Evaluation of the \textit{EEATC} on the stationary deployment of LCS is covered in Sec. \ref{sec:eval_stationary} and on mobile deployments in Sec. \ref{sec:eval_mobile}. The future scope and conclusions are covered in Sec. \ref{sec:conc_future}.

\section{ Estimated Error Augmented Two-phase Calibration (\textit{EEATC}) process} \label{sec:Calibration_process}
\begin{figure}[hbt!]
    \centerline{\includegraphics[width=0.75\columnwidth]{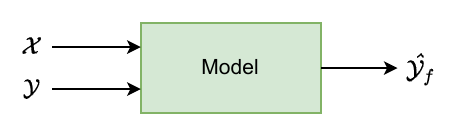}}
    \caption{Block diagram representation of single-phase calibration}
    \label{fig:single_phase_calibration}
\end{figure}

In general, the LCS are calibrated with a single-phase calibration approach shown in Fig. \ref{fig:single_phase_calibration}. In this approach, a model is trained with the LCS data set ($\mathcal{X}$) against the reference station observations ($\mathcal{Y}$), such that the root mean squared error between the sensor output ($\mathcal{S}$) and $\mathcal{Y}$ is minimum. Note that $\mathcal{X}$ is an LCS data set that includes $\mathcal{S}$ and other influencing covariates such as temperature ($T$) and relative humidity ($Rh$). The model in Fig. \ref{fig:single_phase_calibration} may be a machine learning or deep learning model depending on the characteristics of the data. Once the model is trained, it can be used to obtain calibrated sensor values ($\mathcal{\hat{Y}}_f$) for the corresponding $\mathcal{S}$

\begin{figure}[hbt!]
    \centerline{\includegraphics[width=\columnwidth]{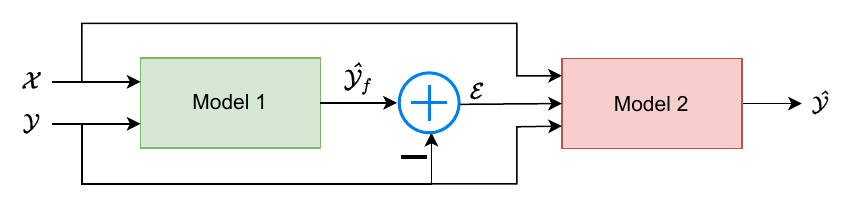}}
    \caption{Block diagram representation of two-phase calibration}
   \label{fig:multiphase_calibration}
\end{figure}

In extension to the single-phase calibration, a two-phase calibration can also be used to calibrate the LCS, which can be implemented by cascading two models as shown in Fig. \ref{fig:multiphase_calibration}. 

In two-phase calibration, the models are cascaded in such a way that the residual error vector ($\mathcal{E}$), the difference between the calibrated output of model 1 ($\mathcal{\hat{Y}}_f$) and the reference instrument values $\mathcal{Y}$, is fed into model 2 along with the input $\mathcal{X}$, that is given in the first phase. In other words, the two-phase calibration can also be treated as a single-phase calibration with augmented input ($\mathcal{X}$ $\vert$ $\mathcal{E}$), where $\mathcal{E}$ is obtained from some process, here that is obtained from other single-phase calibration. Augmentation of $\mathcal{E}$ with $\mathcal{X}$ boosts the performance of the calibration model in the second phase. 

\subsubsection{Importance of error vector $\mathcal{E}$ in boosting the performance}

The calibration of LCS is a regression problem where the calibrated model returns a value for each sample given to the model, and it can be treated as a point prediction. Behind the point prediction, there is always a distribution formed by the model equation and the dimension of the distribution depends on the number of features the model is trained on. 

\begin{figure}[hbt!] 
    \centerline{\includegraphics[width=\columnwidth]{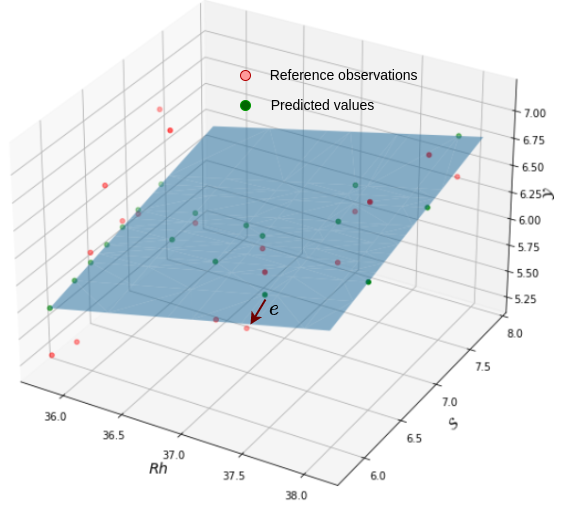}}
    \caption{3D view of the calibration space formed by MLR equation with $S$ and $Rh$}
   \label{fig:3d_plot}
\end{figure}

For instance, in Fig. \ref{fig:single_phase_calibration}, an MLR model is trained for the calibration of LCS with two features $\mathcal{S}$ and $Rh$, then post the training MLR predicts the values ($\mathcal{\hat{Y}}_f$) corresponding to the value of $\mathcal{S}$ and $Rh$ from the posterior distribution formed by the calibration equation in \ref{eq:calibra_equ_first} with calibration coefficients $\beta _1$ and $\beta_2$. The blue mesh in Fig. \ref{fig:3d_plot} is the plane formed by the calibration equation \ref{eq:calibra_equ_first}, and the predicted values ($\mathcal{\hat{Y}}_f$) lie on the plane. However, the predicted values deviate from the reference station observations ($\mathcal{Y}$), which are the residual errors ($e$) in the model predictions shown in Fig. \ref{fig:3d_plot}, and the collection of all such errors forms the residual error vector $\mathcal{E}$ ($\mathcal{E}= \mathcal{\hat{Y}}_f - \mathcal{Y})$. The $\mathcal{E}$ has an evident relation with $\mathcal{Y}$ and can express the uncertainty in the model predictions. The uncertainty in the model predictions can be estimated by analyzing the distribution formed by the $\mathcal{E}$
\begin{equation} \label{eq:calibra_equ_first}
    \mathcal{\hat{Y}}_f= \beta _1 (\mathcal{S}) + \beta _2 (Rh)
\end{equation}

Therefore, including $\mathcal{E}$ in the input to the second phase gives the scope for model 2 to learn the distribution of $\mathcal{E}$ along with $S$ and $Rh$ so that the second phase produces better predictions. Implicitly, instead of conditioning on two variables $S$ and $Rh$ in the single-phase calibration, here in the two-phase calibration, we are conditioning on three variables $\mathcal{E}$, $\mathcal{S}$ and $Rh$, which gives a better estimate of $\mathcal{\hat{Y}}_f$ than the former one. 

Then the question arises as to how to obtain  $\mathcal{E}$. One way is subtracting the calibrated output from the first phase with reference station observations \cite{Lin2018}. Obtaining  $\mathcal{E}$ in this way is possible only in training. However, in testing or remote real-time deployments, the LCS are inaccessible to the reference station observations, which renders the impossibility of obtaining $\mathcal{E}$ in practical applications. Now the only possible way is to use its estimate ($\mathcal{\hat{E}}$) instead of directly using $\mathcal{E}$ in testing or real-time deployments.

\subsubsection{Estimation of $\mathcal{E}$ using nannyML}

Estimation of $\mathcal{E}$ in testing after the first phase can be hypothesised to estimate first-phase model performance without the ground truth or reference station observations. NannyML is a recent development in ML, where an extra ML model is trained on the absolute loss or error of the monitored model \cite{nannyML}. By doing so, the extra ML model can be used to estimate the loss of the monitored model in the lack of ground truth observations. Here, the extra ML model is called the Nanny model, and the monitored model in two-phase calibration is the model at the first phase.  NannyML offers Confidence-based Performance Estimation (CBPE) for classification problems and  Direct Loss Estimation (DLE) for regression problems. Since the calibration of LCS is a regression problem, we have adopted the DLE to estimate the $\mathcal{E}$ after the first phase.

 \begin{figure}[hbt!]
\centerline{\includegraphics[width=\columnwidth]{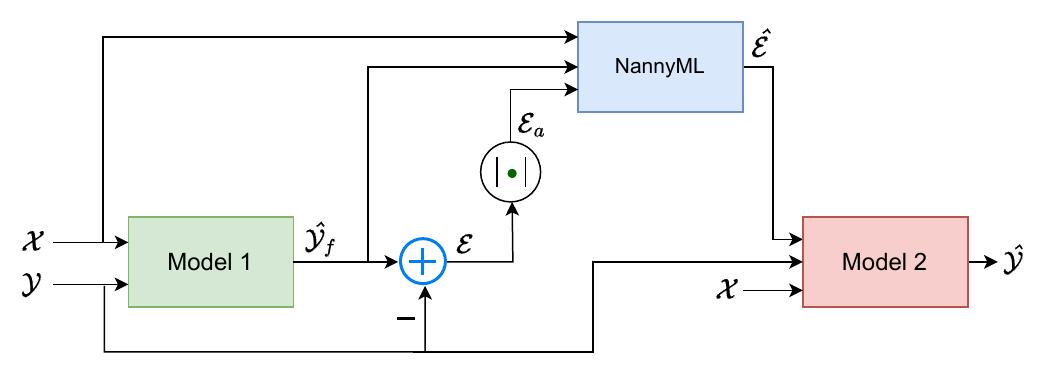}}
    \caption{Block diagram representation of \textit{EEATC} calibration approach}
   \label{fig:EATC}
\end{figure}

Therefore a NannyML model is placed in between the models of the two-phase calibration shown in Fig \ref{fig:multiphase_calibration} such that the $\mathcal{\hat{E}}$ is fed into the second phase instead of $\mathcal{E}$. This is a novel way of implementing two-phase calibration for LCS, and we call this Estimated Error augmented two-phase calibration (\textit{EEATC}). 

The implementation of \textit{EEATC} is shown in Fig. \ref{fig:EATC}. Let’s denote the model in the first phase as $\mathcal{F}$ and 
 the nanny model as $\mathcal{H}$. We are interested in estimating the mean absolute error of $\mathcal{F}$ for which reference station observations  are unavailable. At first, $\mathcal{F}$ is trained with data ($\mathcal{X}$) against $\mathcal{Y}$, then it is used to predict the targets, $\mathcal{\hat{Y}}_f$ as shown in the equation \ref{eq:nanny_predi}
\begin{equation}\label{eq:nanny_predi}
    \mathcal{\hat{Y}}_f = \mathcal{F}(\mathcal{X})
\end{equation}
Then the $\mathcal{H}$ is trained with $\mathcal{X}$ and $\mathcal{\hat{Y}}_f$ against the  absolute error vector ($\mathcal{E}_a$) formed between $\mathcal{\hat{Y}}_f$ and $\mathcal{Y}$ shown in equation \ref{eq:err_vector}
\begin{equation}\label{eq:err_vector}
      \mathcal{E}_a = |\mathcal{Y}-\mathcal{\hat{Y}}_f|
\end{equation}
The trained $\mathcal{H}$ can be used to obtain $\mathcal{\hat{E}}$ vector in real-time without the ground truth observations as shown in equation \ref{eq:err_estimate}
\begin{equation} \label{eq:err_estimate}
    \mathcal{\hat{E}} = \mathcal{F}(\mathcal{X}, \mathcal{\hat{Y}}_f) 
\end{equation}
Where $\mathcal{\hat{E}}$ is the estimated  error vector by nannyML. 
Finally, the model in the second phase leverages $\mathcal{\hat{E}}$ to produce better calibration results compared to the existing approaches.
\subsubsection{When does the performance of the single-phase calibration tend to equal with \textit{EEATC}} \label{subsubsec:eeatc_to_single}

\begin{figure}[hbt!]
\centerline{\includegraphics[width=\columnwidth]{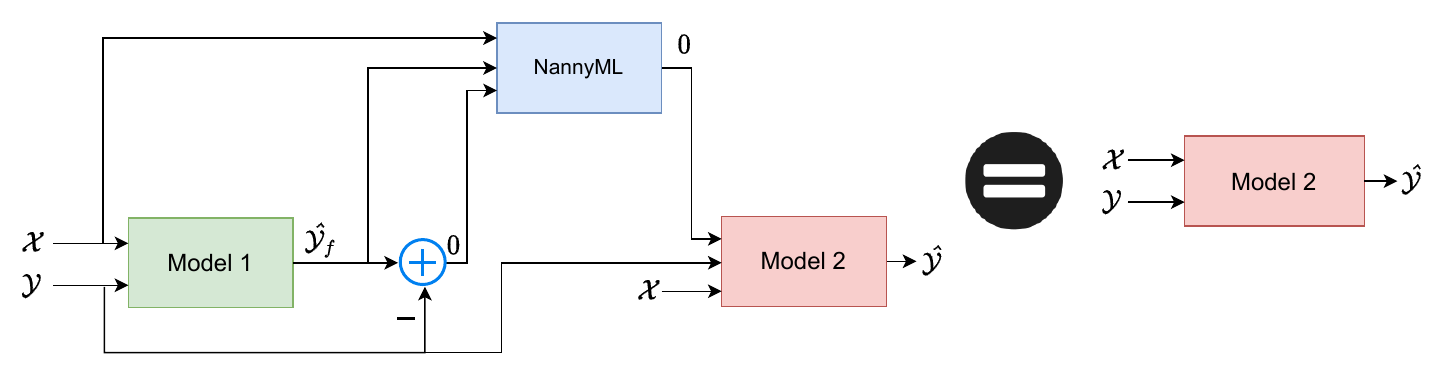}}
    \caption{\textit{EEATC} with $\mathcal{E}$ is a zero vector}
   \label{fig:eeatc_with_e0}
\end{figure}

As we discussed, the two-phase calibration can be represented as single-phase calibration by including additional input $\mathcal{E}$ in the calibration features. When $\mathcal{E}$ is zero, that means the errors between the reference station observations and the calibrated values at the first phase are zero, which implies that the LCS are 100\% calibrated in the first phase itself; if so it doesn't require the second phase calibration. Then the calibration in equation (\ref{eq:2nd_phase_equ}) becomes the calibration equation in (\ref{eq:calibra_equ_first}) provided $\mathcal{E}$ is zero vector. If $\mathcal{E}$ is zero, obviously $\mathcal{\hat{E}}$ is zero, then \textit{EEATC} approach converts to single-phase calibration with model 2, the model in the second phase as shown in Fig. \ref{fig:eeatc_with_e0}. This can happen only in theory, which is practically impossible. 
\begin{equation}\label{eq:2nd_phase_equ} 
    \mathcal{\hat{Y}}= \beta _1 (\mathcal{S}) + \beta _2 (Rh) + \beta _3 (\mathcal{\hat{E}})
    \end{equation}

\subsection{Practical implementation of \textit{EEATC}}
\begin{figure}[hbt!]
    \centerline{\includegraphics[width=\columnwidth]{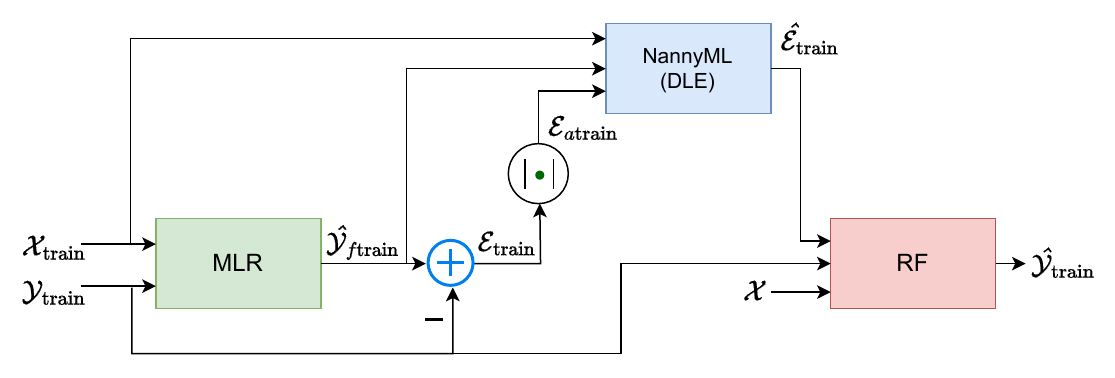}}
    \caption{Practical implementation of \textit{EEATC} (in training) with MLR in the first phase and RF in the second phase}
   \label{fig:twophase_mlr_rf}
\end{figure}

We have implemented the \textit{EEATC} by adopting the MLR model in the first phase and the RF model in the second phase, as shown in Fig.  \ref{fig:twophase_mlr_rf}.  The RF model is a machine learning algorithm that works by constructing an ensemble of decision trees utilizing the training data \cite{Breiman_rf}, and the MLR is a regression model that works by fitting a linear function between the input and target in the training data. The MLR model, in the first phase, learns the linear properties and the RF model, in the second phase, learns the non-linear properties so that the \textit{EEATC} can capture both the properties of data.

We adopted the RF and MLR models from the Scikit-learn library and exhaustively evaluated them for different parameters of the models. The information regarding the parameters is provided in the code in the GitHub link in the resources section. 
\newline \textbf{Training}
We have divided the data sets as training data ($\mathcal{X}_\text{train}$, $\mathcal{Y}_\text{train}$) and testing data ($\mathcal{X}_\text{test}$, $\mathcal{Y}_\text{test}$) with the split of 75\% and 25\% respectively, the standard train test split percentage.  We first trained the MLR model with $\mathcal{X}_\text{train}$ such that loss in equation (\ref{eq:loss}), that is the mean square error (\textbf{MSE}) between the predicted sensors values ($\mathcal{\hat{Y}}_{f\text{train}}$) and the reference instrument values ($\mathcal{Y}_\text{train}$) is minimised. 
     \begin{equation} \label{eq:loss}
     \text{loss}_{\text{MLR}} = \frac{1}{P}\sum_{i=1}^{P}\text{\textbf{MSE}}(\hat{y}^{i}_{f\text{train}}-y^i_\text{train})
\end{equation}
Where $\mathcal{\hat{Y}}_{f\text{train}}$ = $\{ \hat{y}^1_{f\text{train}}, \hat{y}^2_{f\text{train}}, .. P\}$, $\mathcal{Y}_\text{train}$ = $\{ y^1_\text{train}, y^2_\text{train}, .. P\}$ and $P$ is the number of samples in training.

Then the trained MLR model is utilized to generate error vector $\mathcal{E}_{atrain}$ shown in equation (\ref{eq:error}), which is the absolute difference between $\mathcal{\hat{Y}}_{f\text{train}}$ and $\mathcal{Y}_\text{train}$. 
  \begin{equation} \label{eq:error}
         \mathcal{\hat{Y}}_{f\text{train}} = \text{MLR}(\mathcal{X}_\text{train}) ; 
          \mathcal{E}_{a\text{train}} = |\mathcal{\hat{Y}}_{f\text{train}} - \mathcal{Y}_\text{train}|
     \end{equation}
Then the NannyML model ($\mathcal{H}$) is trained with $\mathcal{X}_\text{{train}}$ and $\mathcal{\hat{Y}}_{f\text{train}}$ against the $\mathcal{E}_{a\text{train}}$ obtained in the first phase to learn the uncertainty in the MLR model predictions which can be 
 used to estimate $\mathcal{\hat{E}}_\text{train}$ as shown in equation (\ref{eq:nanny_ml})
\begin{equation}\label{eq:nanny_ml}
    \mathcal{\hat{E}}_\text{train} = \mathcal{H}(\mathcal{X}_\text{train}, \mathcal{\hat{Y}}_{f\text{train}})
\end{equation}
Finally, the model in the second phase, RF, is trained by augmenting the $\mathcal{X}_\text{train}$ with $\mathcal{\hat{E}}_\text{train}$, to minimize the loss in equation (\ref{eq:loss_second_phase}) against $\mathcal{Y}$. Where K is the number of samples in testing. Therefore, P+K = N, the total number of samples in the data sets
\begin{equation} \label{eq:loss_second_phase}
        \text{loss}_{\text{RF}} =  \frac{1}{K}\sum_{i=1}^{K}\text{\textbf{MSE}}(\mathcal{\hat{Y}}^{i}_\text{train}-\mathcal{Y}^i_\text{train})
    \end{equation}
\newline \textbf{Testing} 
\begin{figure}[hbt!]
    \centerline{\includegraphics[width=\columnwidth]{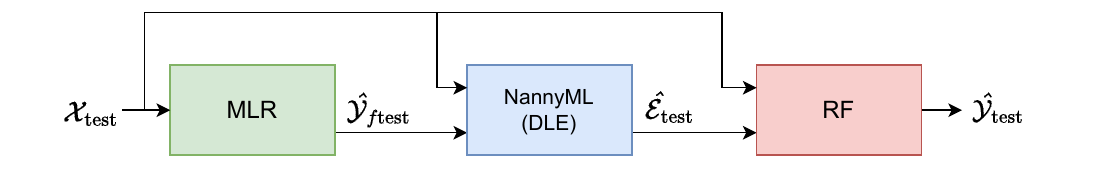}}
    \caption{Testing implementation of \textit{EEATC}  with MLR in the first phase and RF in the second phase}
   \label{fig:twophase_mlr_rf_test}
\end{figure}

Testing is straightforward as shown in Fig. \ref{fig:twophase_mlr_rf_test}, where we fed the test data, $\mathcal{X}_\text{test}$, into the MVLR model and then the predicted $\mathcal{\hat{Y}}_{f\text{test}}$ and $\mathcal{X}_\text{train}$ is passed to nannyML to estimate the error vector, $\mathcal{\hat{E}}_\text{test}$. Once the $\mathcal{\hat{E}}_\text{test}$ is obtained, it is then augmented with $\mathcal{X}_\text{test}$, which is fed to the RF model. Finally, the RF model gives the calibrated values corresponding to the input features values, shown in equation (\ref{eq:output_rf})
\begin{equation} \label{eq:output_rf}
    \mathcal{\hat{Y}} = \text{RF}(\mathcal{X}_\text{test}, \mathcal{\hat{E}}_\text{test})
\end{equation}
Algorithm \ref{alg:implementatino_twophase} illustrates the step by step procedure of practical implementation of \textit{EETAC}

\begin{algorithm}
\caption{Implementation of \textit{EEATC} calibration}\label{alg:implementatino_twophase}
\textbf{Training} \\
\hspace*{\algorithmicindent} \textbf{Input}   $\mathcal{X}_\text{train}, \mathcal{Y}_\text{train}$  \\
    \hspace*{\algorithmicindent} \textbf{Output} $\mathcal{\hat{Y}}_\text{train}$ 
\begin{algorithmic}[1]
\State Train \textbf{MLR} with $\mathcal{X}_\text{train}$ against $\mathcal{Y}_\text{train}$
\State $\mathcal{\hat{Y}}_{f\text{train}}$ = \textbf{MLR}($\mathcal{X}_\text{train}$)
\State $\mathcal{E}_{a\text{train}} = |\mathcal{\hat{Y}}_{f\text{train}}-\mathcal{Y}_\text{train}|$
\State Train \textbf{nannyML} with $\mathcal{X}_\text{train}$, $\mathcal{\hat{Y}}_{f\text{train}}$ against $\mathcal{E}_{a\text{train}}$
\State $\mathcal{\hat{E}}_\text{train} = \text{\textbf{nannyML}}(X_\text{train}, \mathcal{\hat{Y}}_{f\text{train}})$
\State Train \textbf{RF} with $\mathcal{X}_\text{train}$, $\mathcal{\hat{E}}_\text{train}$ against $\mathcal{Y}_\text{train}$ 
\State $\mathcal{\hat{Y}}_\text{train} = \text{\textbf{RF}}(\mathcal{X}_\text{train}, \mathcal{\hat{E}}_\text{train})$
\end{algorithmic}
\textbf{Testing or real-time deployments} \\
\hspace*{\algorithmicindent} \textbf{Input}   $\mathcal{X}_\text{test} $\\
    \hspace*{\algorithmicindent} \textbf{Output} $\mathcal{\hat{Y}}_{test}$, the final calibrated values of $\mathcal{S}$ in real-time applications
\begin{algorithmic}[1]
     \State $\mathcal{\hat{Y}}_{f\text{test}} = \text{\textbf{MLR}}(\mathcal{X}_\text{test})$
     \State $\mathcal{\hat{E}}_\text{test} = \text{\textbf{nannyML}}(\mathcal{X}_\text{test}, \mathcal{\hat{Y}}_\text{test} )$
     \State $\mathcal{\hat{Y}}_{test} = \text{\textbf{RF}}(X_\text{test}, \mathcal{\hat{E}}_\text{test})$
\end{algorithmic}

\end{algorithm}

\subsection{Metrics}
We considered the root mean square error (RMSE) in equation \eqref{eq_rmse} and coefficient of determination ($\text{R}^2$) in equation \eqref{eq_r2} to evaluate the performance of the calibration models. $R^2$  describes the
proportionality of variability between independent and dependent variables of the model. $RMSE$  indicates the deviation between the predicted values and reference station observations.
\begin{equation} 
 \text{RMSE} (\mathcal{\hat{Y}}, \mathcal{Y})=\sqrt{\frac{1}{N}\sum_{j=1}^{N}{(\hat{y}_j-y_j)^2} }
 \label{eq_rmse}
\end{equation}
Here, $\mathcal{\hat{Y}} = \{\hat{y_1}, \hat{y_2}, ... \}_{N}$ are the calibrated sensor values, $\mathcal{Y} = \{y_1, y_2, ....\}_{N} $ are the reference instrument values and $N$ is the number of samples.
\begin{equation}
                 \text{R}^2 = 1-\frac{SS_{res}}{SS_{tot}} 
        \label{eq_r2}
\end{equation}
where 
\begin{equation*}
    SS_{tot} = \sum_{j=1}^{N}{(y_j- \overline{\mathcal{Y}})^2} 
\end{equation*}
\vspace{-0.22cm}
\begin{equation*}  
    {SS_{res}} = \sum_{j=1}^{N}{(y_j-\hat{y}_j)^2} 
\end{equation*}
\vspace{-0.4cm}
\begin{equation*}
    \overline{\mathcal{Y}} = \frac{1}{N}\sum_{i=1}^{N}{y_j}
\end{equation*}

\subsection{Notations at a glance}
\begin{itemize}
    \item  $\mathcal{S}$, $T$ and $Rh$ are the columns vectors representing the LCS values, temperature and relative humidity, respectively, and $\mathcal{X}$ is the data set formed by these vectors after pre-processing. 
    \item $\mathcal{Y}$ is a column vector that represents the  time corresponding reference instrument or reference station observations to the values in $\mathcal{X}$
    \item $\mathcal{\hat{Y}}_f$ and $\mathcal{\hat{Y}}$ are the calibrated values in the first and second phases, respectively.
    \item $\mathcal{E}$ is the residual error vector, and $\mathcal{\hat{E}}$ is the estimated error vector after first phase.
    \item train and test in the subscripts  indicate training and testing 
\end{itemize}
\section{Evaluation of the \textit{EEATC} approach for stationary AQM data sets} \label{sec:eval_stationary}

\begin{table*}[h]
\caption{Evaluation of \textit{EEATC} approach for different sensors in the CAIRSENSE data set and comparison with the well-known calibration models such as MLR and RF. All the approaches are tested with different combinations of input features, and the outperforming one is indicated with *.}
\label{tab:comprison_models_stationary}
    \begin{subtable}[t]{0.5\linewidth}
    \caption{OPCPMF}
        \begin{tabular}{{|l|p{2cm}|cc|cc|}}
            \hline
            \multirow{3}{*}{Model} &\multirow{3}{*}{\shortstack{Variables \\ considered in the \\ calibration process}}  &\multirow{2}{*}{$\text{R}^2$} & & \multirow{2}{*}{RMSE} & \\
            & & \multicolumn{2}{c|}{} & \multicolumn{2}{c|}{}\\ \cline{3-6}
            &    & Training    & Testing  & Training & Testing\\
            \hline
\multirow{4}{*}{MLR}   & $\mathcal{S}$ & 0.16 & 0.16   & 0.92 & 0.93 \\
   &  $\mathcal{S}$, $T$ & 0.21   & 0.21  & 0.89 & 0.91\\
 &  $\mathcal{S}$, $Rh$  &  0.25   &0.25  & 0.86 &0.88 \\
   &  $\mathcal{S}$, $T$, $Rh$  & 0.26   &0.25  & 0.86 & 0.88\\
\hline
\multirow{4}{*}{RF}   & $\mathcal{S}$ & 0.29 & 0.27   & 0.84 & 0.87 \\
 &  $\mathcal{S}$, $T$  & 0.48   & 0.45 &0.72 &0.75 \\
  &  $\mathcal{S}$, $Rh$ & 0.46   & 0.44  & 0.74 & 0.76\\
 &  $\mathcal{S}$, $T$, $Rh$  & 0.55    & 0.53   & 0.67  & 0.7  \\
\hline
\multirow{4}{*}{\textit{EEATC}}   & $\mathcal{S}$ & 0.3 & 0.27   & 0.84 & 0.86 \\
 &  $\mathcal{S}$, $T$  & 0.58   & 0.47 &0.65 &0.74 \\
  &  $\mathcal{S}$, $Rh$ & 0.55   & 0.46 & 0.67 & 0.75\\
 &  $\mathcal{S}$, $T$, $Rh$   & 0.76*    & 0.67*   & 0.49*  & 0.58*  \\
 \hline
        \end{tabular}
    \end{subtable}
     \hfil
     \begin{subtable}[t]{0.5\linewidth}
    \caption{Speck}
        \begin{tabular}{{|l|p{2cm}|cc|cc|}}
            \hline
            \multirow{3}{*}{Model} &\multirow{3}{*}{\shortstack{Variables \\ considered in the \\ calibration process}}  &\multirow{2}{*}{$\text{R}^2$} & & \multirow{2}{*}{RMSE} & \\
            & & \multicolumn{2}{c|}{} & \multicolumn{2}{c|}{}\\ \cline{3-6}
            &    & Training    & Testing  & Training & Testing\\
            \hline
\multirow{4}{*}{MLR}   & $\mathcal{S}$ & 0.14 & 0.14   & 0.93 & 0.95 \\
   &  $\mathcal{S}$, $T$ & 0.37   & 0.37  & 0.79 & 0.82\\
 &  $\mathcal{S}$, $Rh$  &  0.37   &0.36  & 0.8 &0.82 \\
   &  $\mathcal{S}$, $T$, $Rh$  & 0.43   &0.42  & 0.75 & 0.78\\
\hline
\multirow{4}{*}{RF}   & $\mathcal{S}$ & 0.15 & 0.15   & 0.92 & 0.95 \\

 &  $\mathcal{S}$, $T$  & 0.53   & 0.52 &0.68 &0.71 \\
  &  $\mathcal{S}$, $Rh$ & 0.43   & 0.43  & 0.75 & 0.78 \\
 &  $\mathcal{S}$, $T$, $Rh$  & 0.61    & 0.6   & 0.62  & 0.65  \\
\hline
\multirow{4}{*}{\textit{EEATC}}   & $\mathcal{S}$  & 0.15   & 0.15 & 0.92 & 0.95 \\

 &  $\mathcal{S}$, $T$  & 0.59   & 0.54 & 0.64 & 0.7 \\
  &  $\mathcal{S}$, $Rh$ & 0.52   & 0.46 & 0.69 & 0.75 \\
 &  $\mathcal{S}$, $T$, $Rh$   & 0.77*    & 0.73*   & 0.48*  & 0.5*  \\
\hline  
        \end{tabular}
    \end{subtable}
    \\[1.5em]
    \begin{subtable}[t]{0.5\linewidth}
    \caption{TZOA}
        \begin{tabular}{{|l|p{2cm}|cc|cc|}}
            \hline
            \multirow{3}{*}{Model} &\multirow{3}{*}{\shortstack{Variables \\ considered in the \\ calibration process}}  &\multirow{2}{*}{$\text{R}^2$} & & \multirow{2}{*}{RMSE} & \\
            & & \multicolumn{2}{c|}{} & \multicolumn{2}{c|}{}\\ \cline{3-6}
            &    & Training    & Testing  & Training & Testing\\
            \hline
\multirow{4}{*}{MLR}   & $\mathcal{S}$ & 0.46 & 0.39   & 0.73 & 0.82 \\
   &  $\mathcal{S}$, $T$ & 0.55   & 0.47  & 0.67 & 0.77 \\
 &  $\mathcal{S}$, $Rh$  &  0.49  &0.41  & 0.72 &0.80 \\
   &  $\mathcal{S}$, $T$, $Rh$  & 0.55   &0.47  & 0.67 & 0.76\\
\hline
\multirow{4}{*}{RF}   & $\mathcal{S}$ & 0.51 & 0.40   & 0.7 & 0.81 \\

 &  $\mathcal{S}$, $T$  & 0.70   & 0.51 &0.55 &0.73 \\
  &  $\mathcal{S}$, $Rh$ & 0.62   & 0.45  & 0.62 & 0.78\\
 &  $\mathcal{S}$, $T$, $Rh$  & 0.69    & 0.52   & 0.56 & 0.73 \\
\hline
\multirow{4}{*}{\textit{EEATC}}   & $\mathcal{S}$  & 0.51   & 0.4 & 0.7 & 0.81 \\

 &  $\mathcal{S}$, $T$  & 0.81   & 0.54 & 0.43 & 0.71 \\
  &  $\mathcal{S}$, $Rh$ & 0.7    & 0.42   & 0.55  & 0.8  \\
 &  $\mathcal{S}$, $T$, $Rh$   & 0.85*    & 0.62*   & 0.38*  & 0.65*  \\
\hline  
        \end{tabular}
    \end{subtable}
    \begin{subtable}[t]{0.5\linewidth}
    \caption{AirAssure}
        \begin{tabular}{{|l|p{2cm}|cc|cc|}}
            \hline
            \multirow{3}{*}{Model} &\multirow{3}{*}{\shortstack{Variables \\ considered in the \\ calibration process}}  &\multirow{2}{*}{$\text{R}^2$} & & \multirow{2}{*}{RMSE} & \\
            & & \multicolumn{2}{c|}{} & \multicolumn{2}{c|}{}\\ \cline{3-6}
            &    & Training    & Testing  & Training & Testing\\
            \hline
\multirow{4}{*}{MLR}   & $\mathcal{S}$ & 0.56 & 0.56   & 0.67 & 0.67 \\
   &  $\mathcal{S}$, $T$ & 0.57   & 0.57  & 0.66 & 0.66\\
 &  $\mathcal{S}$, $Rh$  &  0.57   &0.58  & 0.65 &0.66 \\
   &  $\mathcal{S}$, $T$, $Rh$  & 0.57   &0.58  & 0.65 & 0.66\\
\hline
\multirow{4}{*}{RF}   & $\mathcal{S}$ & 0.67 & 0.63   & 0.6 & 0.62 \\

 &  $\mathcal{S}$, $T$  & 0.69  & 0.68 &0.55 &0.57 \\
  &  $\mathcal{S}$, $Rh$ & 0.7   & 0.68  & 0.55 & 0.57\\
 &  $\mathcal{S}$, $T$, $Rh$  & 0.73    & 0.71   & 0.52 & 0.55 \\
\hline
\multirow{4}{*}{\textit{EEATC}}   & $\mathcal{S}$  & 0.64   & 0.63 & 0.6 & 0.62 \\

 &  $\mathcal{S}$, $T$  & 0.73   & 0.69 & 0.6 & 0.62 \\
  &  $\mathcal{S}$, $Rh$ & 0.74   & 0.69 & 0.51 & 0.57 \\
 &  $\mathcal{S}$, $T$, $Rh$   & 0.80*    & 0.75*   & 0.44*  & 0.5*  \\
\hline  
        \end{tabular}
    \end{subtable}
     \\[1.5em]
    \begin{subtable}[t]{0.5\linewidth}
    \caption{Shinyei}
        \begin{tabular}{{|l|p{2cm}|cc|cc|}}
            \hline
            \multirow{3}{*}{Model} &\multirow{3}{*}{\shortstack{Variables \\ considered in the \\ calibration process}}  &\multirow{2}{*}{$\text{R}^2$} & & \multirow{2}{*}{RMSE} & \\
            & & \multicolumn{2}{c|}{} & \multicolumn{2}{c|}{}\\ \cline{3-6}
            &    & Training    & Testing  & Training & Testing\\
            \hline
\multirow{4}{*}{MLR}   & $\mathcal{S}$ & 0.48 & 0.46   & 0.72 & 0.75 \\
   &  $\mathcal{S}$, $T$ & 0.55   & 0.53  & 0.67 & 0.7 \\
 &  $\mathcal{S}$, $Rh$  &  0.54   &0.52  & 0.68 &0.70 \\
   &  $\mathcal{S}$, $T$, $Rh$  & 0.56   &0.54  & 0.66 & 0.69 \\
\hline
\multirow{4}{*}{RF}   & $\mathcal{S}$ & 0.63 & 0.63   & 0.61 & 0.62 \\

 &  $\mathcal{S}$, $T$  & 0.76  & 0.76 &0.49 &0.49 \\
  &  $\mathcal{S}$, $Rh$ & 0.73   & 0.74  & 0.52 & 0.52\\
 &  $\mathcal{S}$, $T$, $Rh$  & 0.78    & 0.78   & 0.47 & 0.47 \\
\hline
\hline
\multirow{4}{*}{\textit{EEATC}}   & $\mathcal{S}$  & 0.63   & 0.63 & 0.61 & 0.62 \\

 &  $\mathcal{S}$, $T$  & 0.79   & 0.76 & 0.46 & 0.50 \\
  &  $\mathcal{S}$, $Rh$ & 0.75   & 0.75 & 0.48 & 0.51 \\
 &  $\mathcal{S}$, $T$, $Rh$   & 0.86*    & 0.83*   & 0.38*  & 0.42* \\
\hline  
        \end{tabular}
    \end{subtable}
    \begin{subtable}[t]{0.5\linewidth}
    \caption{Aeroqual}
        \begin{tabular}{{|l|p{2cm}|cc|cc|}}
            \hline
            \multirow{3}{*}{Model} &\multirow{3}{*}{\shortstack{Variables \\ considered in the \\ calibration process}}  &\multirow{2}{*}{$\text{R}^2$} & & \multirow{2}{*}{RMSE} & \\
            & & \multicolumn{2}{c|}{} & \multicolumn{2}{c|}{}\\ \cline{3-6}
            &    & Training    & Testing  & Training & Testing\\
           \hline
\multirow{4}{*}{MLR}   & $\mathcal{S}$ & 0.90 & 0.90   & 0.31 & 0.31 \\
   &  $\mathcal{S}$, $T$ & 0.90   & 0.90  & 0.31 & 0.31\\
 &  $\mathcal{S}$, $Rh$  &  0.90   &0.90  & 0.31 &0.31 \\
   &  $\mathcal{S}$, $T$, $Rh$  & 0.90   &0.90  & 0.31 & 0.31\\
\hline
\multirow{4}{*}{RF}   & $\mathcal{S}$ & 0.93 & 0.93   & 0.26 & 0.26 \\

 &  $\mathcal{S}$, $T$  & 0.94  & 0.94 &0.23 &0.23 \\
  &  $\mathcal{S}$, $Rh$ & 0.94   & 0.94  & 0.24 & 0.24\\
 &  $\mathcal{S}$, $T$, $Rh$  & 0.95    & 0.95   & 0.23 & 0.23 \\
\hline
\multirow{4}{*}{\textit{EEATC}}   & $\mathcal{S}$  & 0.93   & 0.93 & 0.26 & 0.26 \\

 &  $\mathcal{S}$, $T$  & 0.95   & 0.94 & 0.23 & 0.24 \\
  &  $\mathcal{S}$, $Rh$ & 0.94   & 0.94 & 0.24 & 0.26 \\
 &  $\mathcal{S}$, $T$, $Rh$   & 0.96*    & 0.95*   & 0.2*  & 0.23*  \\
 \hline
        \end{tabular}
    \end{subtable}
\end{table*}
In general, LCS for AQM are divided into metal oxide ($MO_x$), electrochemical ($EC$) and optical sensors ($OS$) based on the principle on which they work. $MO_x$ and $EC$ are predominant in gas pollutants monitoring and $OS$ sensors are known for PM monitoring. Relative humidity($Rh$) and temperature ($T$) are the prime environmental factors influencing the accuracy of $MO_x$ and $EC$ sensors. These factors influence the resistance of the heating element under the metal oxide surface in $MO_x$ sensors; thus, changing its sensitivity causes inaccurate measurements \cite{Aurora_correction, Masson_correction}. In the case of $EC$ sensors, $T$ changes the rate of the chemical reaction and moisture content depletes the sensors' electrodes, thus changing the response of the sensors \cite{Popoola_correction, Wei_correction}. In addition, the gas sensors suffer from low selectivity, which means they are cross-sensitive to other gas pollutants present in the atmosphere \cite{maag_predep, Liu_gassensors}.

PM sensors work on optical principles severely affected by higher moisture content since particles experience hygroscopic growth at higher $Rh$, where the  tiny particles floating around tend to stick together more because of the water vapour \cite{wang_pm_rheffect}. The hygroscopic growth factor can be calculated as the ratio of PM at a given humidity and temperature to that of the PM at the temperature and humidity where the reference instrument is reporting the data. Carl et al. \cite{Malings_correction} used the same method to correct the hygroscopic growth effect on particulate matter measured by low-cost sensors working on the light scattering principle. Later when they used the data-driven approaches, they used temperature and humidity as explanatory variables of the model to calibrate the same sensors since the hygroscopic growth factor accounts for both temperature and humidity, and the ML algorithms take care of the complex relationship among the variables. A similar procedure is followed by Zheng et al. \cite{Zheng_correction}, Badural et al. \cite{Badura_sl_ml_ann}, Janani et al. \cite{Janani_temp} to calibrate the low-cost sensors works on light scattering principle. 

In order to test \textit{EEATC} on different LCS in stationary deployments, we choose Community Air Sensor Network (CAIRSENSE) data set. The CAIRSENSE data set is an open repository developed by Feinberg et al. \cite{cairsense_dataset} for LCS, and it is available at the US Environmental Protection Agency (USEPA) data website \cite{us_epa}. The data set contains different LCS measurements ($\mathcal{S}$) and the time corresponding $T$, $Rh$ and $\mathcal{Y}$ values. The measurements are taken from the Colorado state at a one-minute resolution for one year. 

The list of sensors involves in testing from the CAIRSENSE data set is as follows, 
\begin{itemize}
    \item Speck -  measures PM - optical principle
    \item Tzoa - Measures PM  - optical principle
    \item Aeroqual - measures $O_3$ - metal oxide principle
    \item Shinyei - measures PM - optical principle
    \item OPCPMF -  measures PM - optical principle
    \item AirAssure - measures PM - optical principle
\end{itemize}

Before applying the calibration procedures, we eliminated the data outliers based on threshold-based filtering. Then we normalized the data with mean and standard deviation normalization, which is then separated into training and testing at 75\% and 25 \%, respectively. All data cleaning, training and testing parameters are available in the code at the GitHub link in the resources section. 

Our approach is data-driven, a general setup of any machine learning-based calibration. The input features on which the model builds are the model's explanatory variables, and these are the parameters on which the performance of the selected sensors gets influenced. Since there exist studies on the effect of meteorology on the performance of LCS, including sensors that work on optical principles, as a cautionary note, we tested all the models with different feature subsets with cross-validation, and the results are presented in table \ref{tab:comprison_models_stationary}. The outperforming combinations of features are indicated with *.

\begin{figure}[hbt!]
     \centering
     \subfloat[\label{fig:eatc_train}]{%
        \includegraphics[width = \columnwidth]{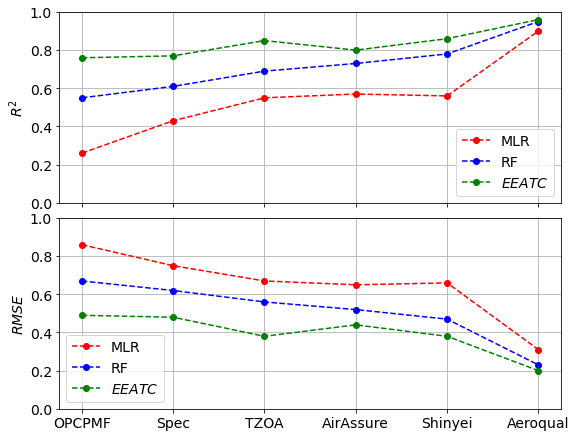}}
        \\
     \subfloat[\label{fig:eatc_test}]{%
         \includegraphics[width = \columnwidth]{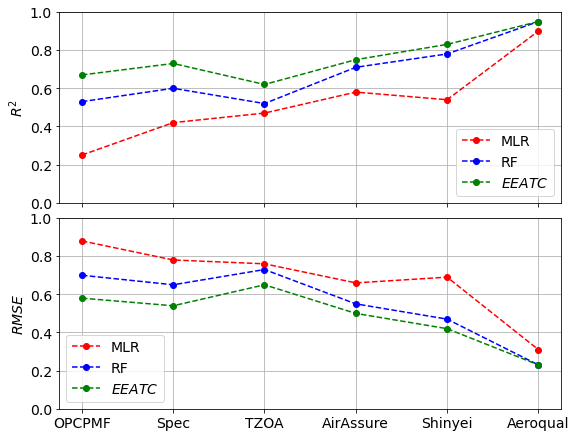}}
         \\
         \caption{comparison of \textit{EEATC} approach with MLR and RF in terms of $R^2$ and $RMSE$ for different sensors (a) in training (b) in testing }
\end{figure} 

The plots drawn for the outperforming combination of features from the table \ref{tab:comprison_models_stationary} in Fig. \ref{fig:eatc_train} and \ref{fig:eatc_test} show the outperformance of the \textit{EEATC} approach compared to the well-known calibration models, such as MLR and RF for all the sensors in training and testing respectively. Therefore, we conclude that the \textit{EEATC} performs better than the MLR and RF approaches in stationary deployments. 

However, in the case of Aeroqual sensors,  $R^2$ of 0.9 and $RMSE$ of 0.31 with the MLR model clearly show the linear relationship among the features utilized in the calibration process. As we discussed in the section \ref{subsubsec:eeatc_to_single}, when the single phase calibration itself can calibrate the LCS, then the performance difference between \textit{EEATC} and single phase approach is less. It can be verified in Fig. \ref{fig:eatc_train} and \ref{fig:eatc_test}.

\section{Evaluation of the \textit{EEATC} approach for mobile AQM data sets} \label{sec:eval_mobile}
\subsection{Mobile monitoring} Due to lack of publicly available data sets on mobile monitoring with LCS, we  conducted a pilot in Chennai (the city of Chennai spans nearly 1000 ${km^2}$), India. For that, we have designed SensurAir, an LCS device which monitors PM.

The core of the SensurAir is the 32-bit microcontroller, STM-32,  by STMicroelectronics, that integrates our LCS. Two LCSs, SDS011 and BME280, and a GSM module, SIM900A, are integrated into the STM-32 on a customized printed circuit board. SDS011 is a well-known LCS \cite{patwardhan_sds011, Rai2017} to monitor PM and works on the light scattering principle, where the intensity of scattered light indicates the PM concentration. The photodiode in SDS011 measures the intensity of scattered light and converts it into an electrical signal which is transformed into particle count and size by a predefined algorithm \cite{Light_scattering}.   

BME280 is a meteorological sensor specially developed for mobile and wearable applications \cite{Tagle_bme280}. It consumes less power and measures temperature ($T$) and relative humidity ($Rh$). Though it is an LCS, it is used as a reference device to measure meteorological parameters in sensor devices since it is known to be very accurate \cite{Sayahi2019}. The GSM module of the SensurAir device transfers sensor data to the central server. A GPS logger captures the vehicle's location during the monitoring period. 
\begin{figure}[hbt!]
    \centerline{\includegraphics[width=0.9\columnwidth, trim = 0cm 2cm 0cm 0cm, clip]{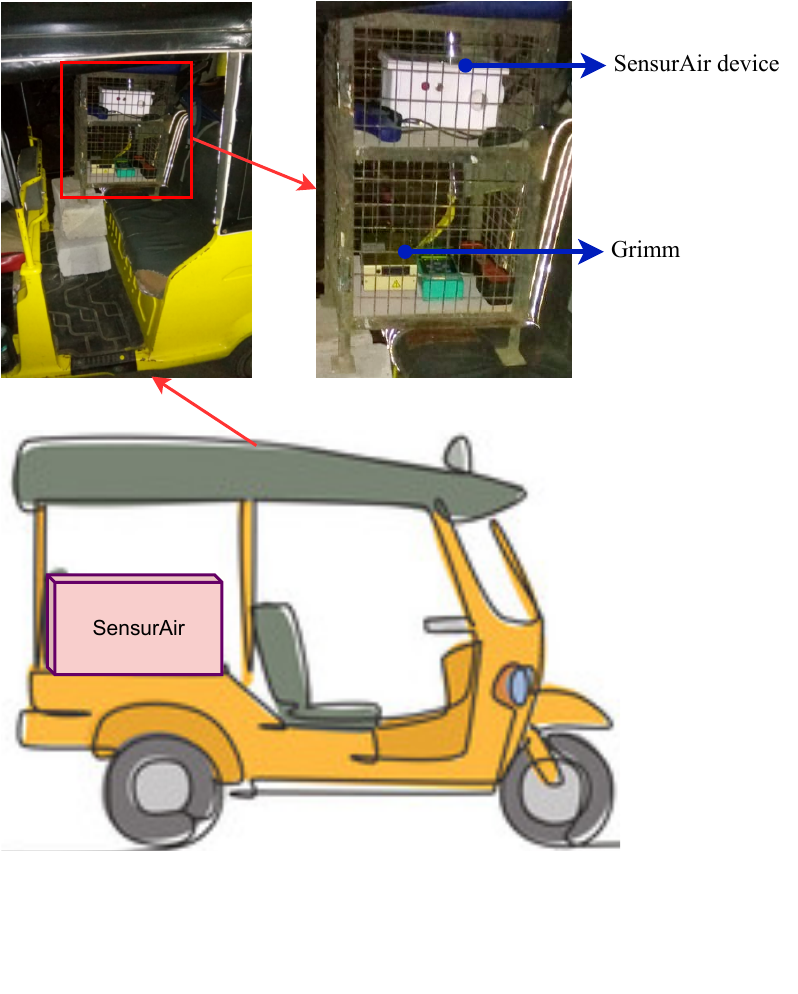}}
         \caption{SensurAir mobile monitoring setup}
        \label{fig:monitoring_setup}
\end{figure}

Grimm, a nationally approved instrument for particulate monitoring, was used as a reference device during the pilot. Grimm continuously samples air through a pump, measures every 6 seconds, and averages over a minute. Our mobile monitoring device travels at an average speed of 5.1  kmph (1.4 m/s), shown in table \ref{tab:statistics}, implying that Grimm captures 60 values within a distance of 84 m and averages them to obtain one measurement reading. Therefore, the performance of Grimm is not affected by vehicle speed in our mobile monitoring. Using the same principle, Wang et al. \cite{wang_mobile} used Grimm in a mobile laboratory to monitor on-road air pollutants during the Beijing 2008 summer Olympics. Elen et al. \cite{elen_cycle} used Grimm on bicycles to monitor PM. At the same time, Grimm and SDS011, the LCS in the SensurAir,  work on the light scattering principle. Therefore we can compare the final results of the calibration without any ambiguity. The widespread use of Grimm motivates its selection as a reference instrument for mobile monitoring.

\begin{figure}[hbt!]
    \centerline{\includegraphics[width=0.75\columnwidth]{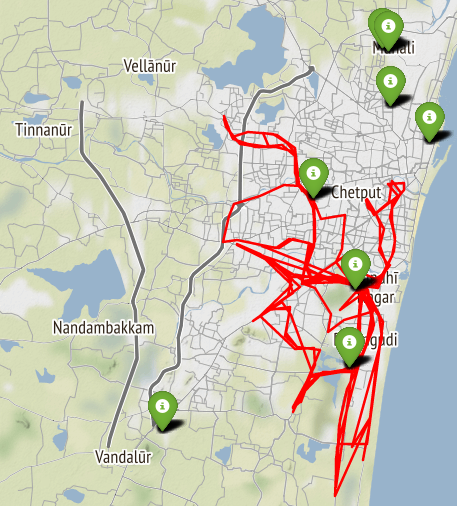}}
    \caption{ Mobile monitoring on Google Maps. The red track is the monitoring route, and the green markers are the reference stations near the route. }. 
    \label{fig:Routemap}
\end{figure}

We have deployed the SensurAir devices and the reference instrument (Grimm) on a rickshaw as shown in Fig. \ref{fig:monitoring_setup} and operate opportunistically in parts of Chennai, India. The route map of mobile monitoring is shown in Figure \ref{fig:Routemap}.  As we mentioned, the vehicle moves opportunistically, and the positions of the vehicle are captured with the GPS module. Whenever there was a dead end in the roads or the end of the monitoring period on that day, we started the vehicle from the new positions. We plot the raw GPS positions without tailoring them on the map, which may be the reason for some sudden jumps on the track shown in Fig.2. The main reason for selecting a rickshaw is that it is open on the sides to the atmosphere and provides unrestricted airflow to the instruments.
Mobile monitoring was conducted over three weeks, in the mornings (usually between 10.30 am and 12.30 pm) and evenings (between 6.30 and 9.30 pm), between September and October 2021. 
\subsection{Measurements}
\begin{figure}[hbt!]
     \centering
     \subfloat[\label{fig:PM25_sensor_values}]{%
        \includegraphics[width = \columnwidth]{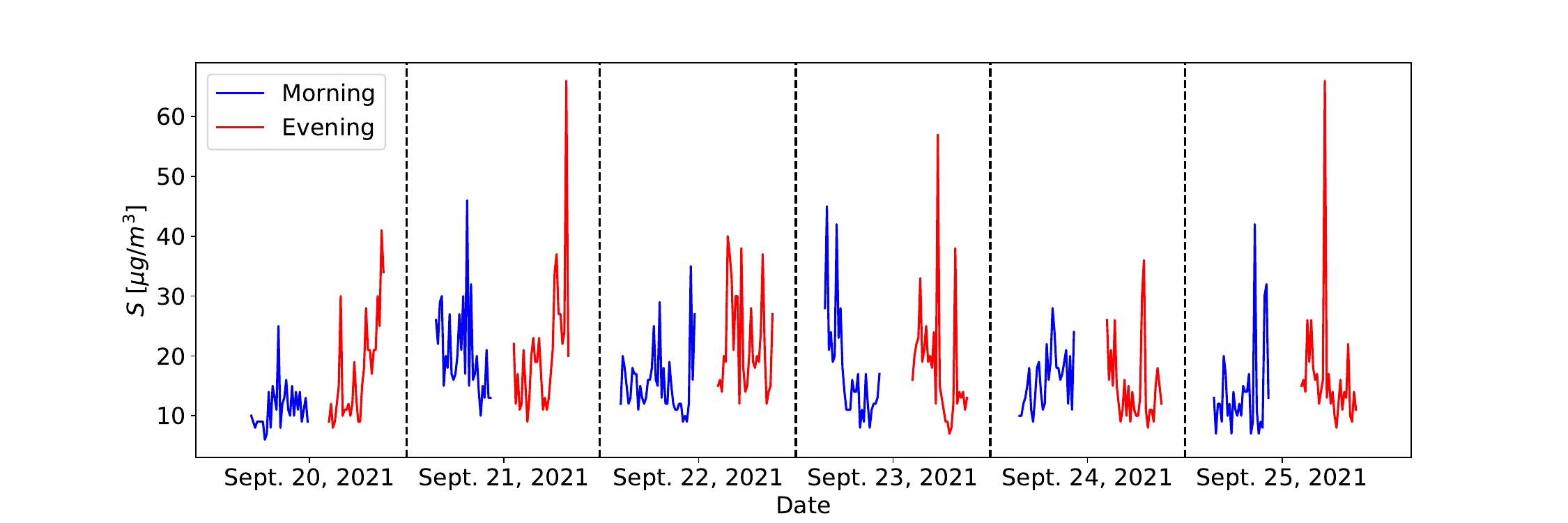}}
        \\
        \vspace{-0.1cm}
     \subfloat[\label{fig:PM25_grimm_values}]{%
         \includegraphics[width = \columnwidth]{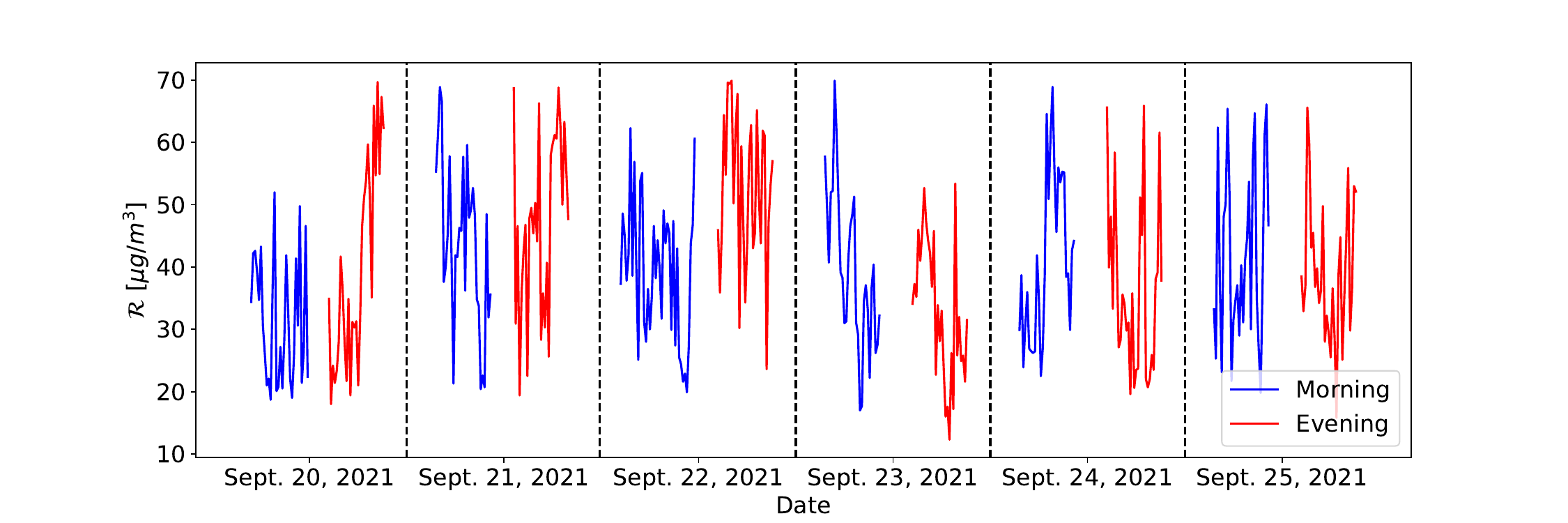}}
         \\
         
         \caption{ $PM_{25}$ values of mobile monitoring for six consecutive days for morning and evening. (a) from LCS   (b) from Grimm}
        \label{fig:pm25sen_pm25grim}
\end{figure}

\begin{figure}[hbt!]
     \centering
     \subfloat[\label{fig:rh}]{%
        \includegraphics[width = \columnwidth]{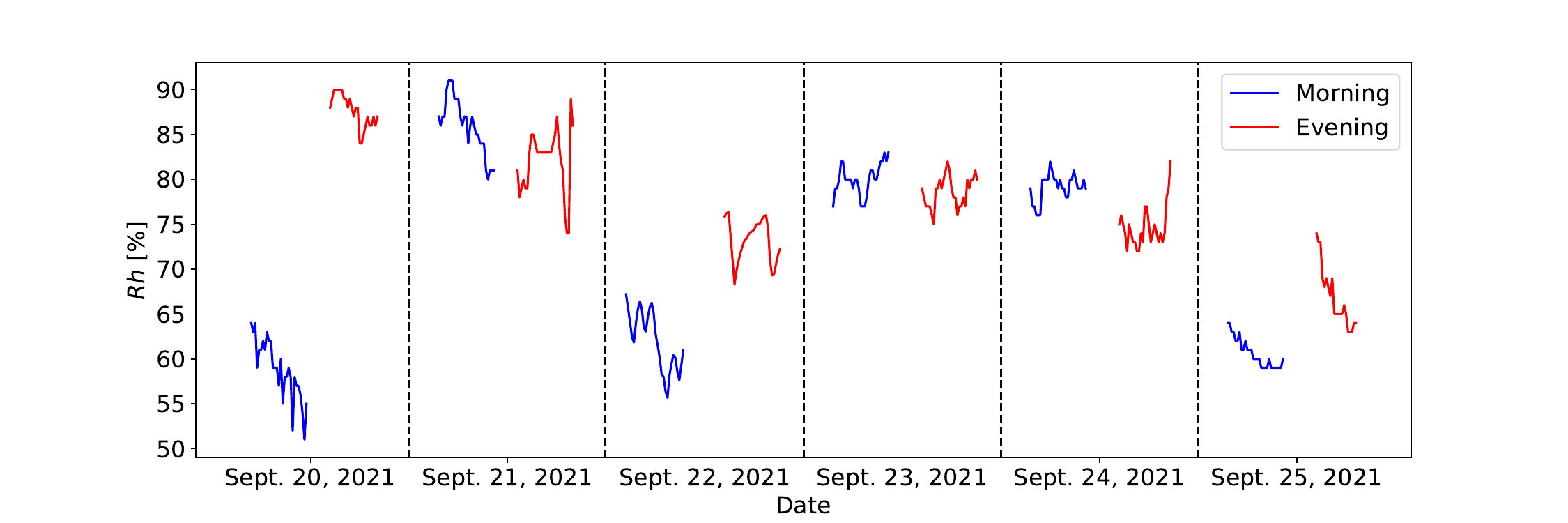}}
        \\
        \vspace{-0.1cm}
     \subfloat[\label{fig:temp}]{%
         \includegraphics[width = \columnwidth]{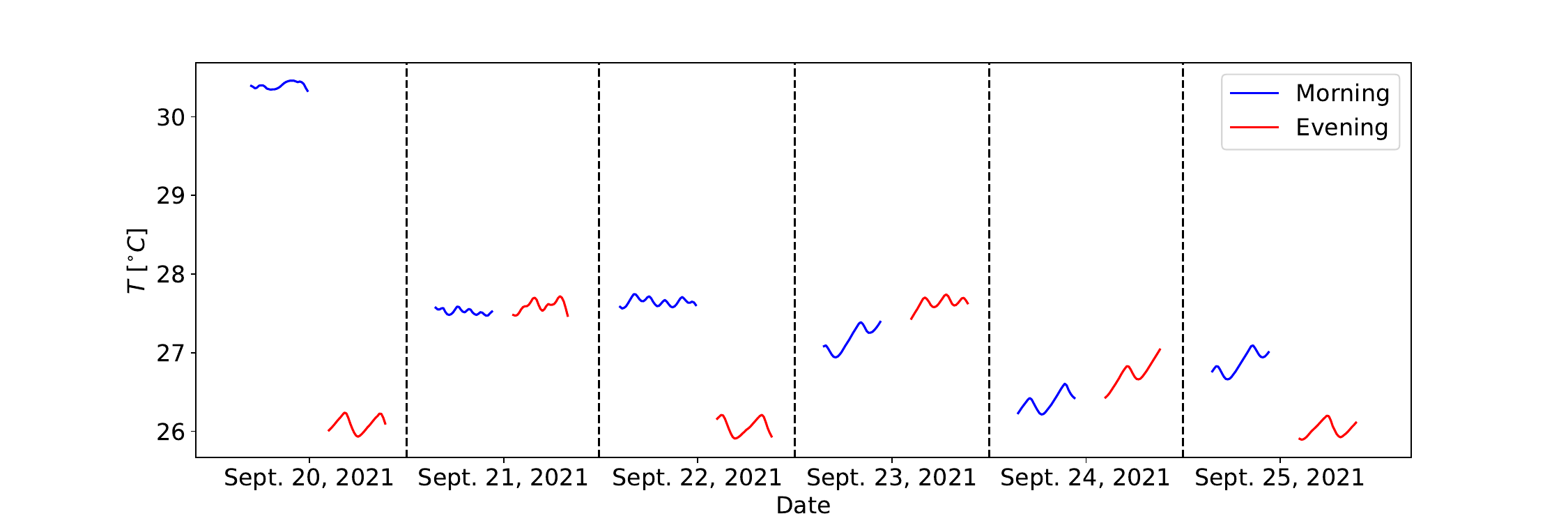}}
         \\
         \caption{ Meteorological measurements of mobile monitoring for six consecutive days for morning and evening. (a) $Rh$ (b)$T$ }
        \label{fig:temp_rh}
\end{figure}

During the pilot, we monitored $PM_{25}$, $PM_{10}$, $T$, and$Rh$ parameters. However, the present work covers the calibration of $PM_{25}$ for the purpose of illustration, and the same procedure outlined holds for $PM_{10}$. The sample plots of $PM_{25}$, $T$ and $Rh$ for six consecutive days in the morning and the evening are shown in Fig. \ref{fig:pm25sen_pm25grim} and \ref{fig:temp_rh} and Table \ref{tab:statistics} illustrates the statistics of these parameters for the entire monitoring period. 

From Fig. \ref{fig:pm25sen_pm25grim}, it can be observed that the $PM_{25}$ concentrations in the evenings are slightly higher than what is observed in the mornings. This may be due to the evening after-office rush hour traffic. Further, $T$ values are higher during the mornings than in the evenings, which is an expected scenario. The variations in $Rh$ are the exact opposite of $T$, which is a general tendency of meteorology, and our measurements follow the same.


\subsection{Data set preparation} \label{subsec:data_sets}
All the measurements are collected at a resolution of one second and then averaged the data over a minute. This averaging process serves two purposes. Firstly, it allows time synchronization with the reference instrument measurements, enabling comparisons between the LCS and reference measurements. Secondly, it helps to smooth out any transient effects in the data caused by the mobility of the vehicle carrying the LCS \cite{Lin2018}. We followed the methodology of Dong et al. \cite{Dong_mobile_ann} to remove the samples of vehicle start and stop positions from the data sets using GPS positions.  Then, we applied a threshold-based filtering process to remove the outliers in the data

In order to address the heterogeneity in the data, we normalized all the parameters using mean and standard deviation normalization. This normalization ensures all the parameters have a  similar scale and distribution, facilitating faster learning for the calibration models.
\begin{table}[hbt!]
\caption{Statistics of the mobile monitoring data.}
\label{tab:statistics}
\begin{tabular}{|c|c|c|c|c|c|}
\hline
parameter & \makecell{$\mathcal{S}$\\ ($\mu g/m^3$)}   & \makecell{$\mathcal{R}$\\ ($\mu g/m^3$)} & $Rh$ ($\%$) & $T$ ($^{\circ} C $)  & \makecell {$\text{VS}$ \\($kmph$)} \\

\hline 
Min      & 4.3      & 7.7       & 51 & 26   & 2.4 \\
Max      & 59          & 69.9           &91    &36.2      &29.4     \\
 Mean         &19.9          &35.9           &71.6    &30.6      &5.1     \\
 Std          &7.8          &12.3           &7.5    &2.6      &2.38  \\
\hline 
\end{tabular}
\end{table}
We then constructed a data set $\phi$ based on the normalized data and it contains three features $\mathcal{S}$, $T$ and $Rh$. It provides the input data for the calibration models and this is the mobile LCS data set contributed to the research community. Therefore, it can be represented in matrix form as follows, where each row of the matrix corresponds to a specific observation or measurement, and each column represents a specific parameter or input. 

\begin{align*}
\phi = \begin{bmatrix}
\mathcal{S} &  T & Rh \\
\end{bmatrix}_{(N \times 3)}
\end{align*}

\subsection{Evaluation} \label{sec:results}
Following the data set preparation, we tested the MLR and RF models with $\phi$. The primary purpose of this testing is to check how these models work with the mobile LCS data, and, at the same time, this helps to compare the performance of the proposed method with these baseline models. We trained the models with 75 \% of the random data from $\phi$ and tested them with the remaining data. To show the impact of each variable on the calibration process, we have trained and tested the models with different combinations of the variables, and the results are presented in Table  \ref{tab:single_phase}.  The best-performing combination is highlighted with $*$.
\begin{table}[hbt!]
\centering
\caption{Performance of various models in the calibration of mobile LCS with Single-phase calibration }
\label{tab:single_phase}
\begin{tabular}{{|l|p{2cm}|cc|cc|}}
\hline
\multirow{3}{*}{Model} &\multirow{3}{*}{\shortstack{Variables \\ considered in the \\ calibration process}}  &\multirow{2}{*}{$\text{R}^2$} & & \multirow{2}{*}{RMSE} & \\
& & \multicolumn{2}{c|}{} & \multicolumn{2}{c|}{}\\ \cline{3-6}
&    & Training    & Testing  & Training & Testing\\
\hline 
\multirow{4}{*}{MLR}   & $\mathcal{S}$ & 0.12 & 0.13   & 0.93 & 0.81 \\
   &  $\mathcal{S}$+$Rh$ & 0.16   & 0.19  & 0.91 & 0.78\\
 &  $\mathcal{S}$ +$T$  &  0.27   &0.27  & 0.85 &0.74 \\
   &  $\mathcal{S}$+$T$+$Rh$ ($\phi$)  & 0.36*   &0.41*  & 0.79* & 0.66 *\\
\hline
\multirow{4}{*}{RF}   & $\mathcal{S}$ & 0.35 & 0.14   & 0.80 & 0.81 \\

 &  $\mathcal{S}$+$Rh$  & 0.47   & 0.44 &0.72 &0.65 \\
  &  $\mathcal{S}$+ $T$ & 0.63   & 0.59  & 0.60 & 0.59\\
 &  $\mathcal{S}$+$T$+$Rh$ ($\phi$)  & 0.68 *   & 0.64 *  & 0.56 * & 0.52 * \\
\hline
\multirow{4}{*}{\textit{EEATC}}   & $\mathcal{S}$ & 0.66 & 0.50   & 0.57 & 0.61 \\

 &  $\mathcal{S}$+$Rh$  & 0.70   & 0.52 &0.54 &0.60 \\
  &  $\mathcal{S}$+ $T$ & 0.85   & 0.78 & 0.38 & 0.40\\
 &  $\mathcal{S}$+$T$+$Rh$ ($\phi$)  & 0.86 *   & 0.82 *  & 0.37 * & 0.37 * \\
\hline  
\end{tabular}
\end{table} 

\begin{figure}[hbt!]
\centerline{\includegraphics[width=\columnwidth]{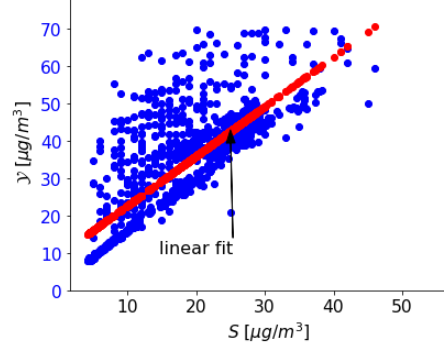}}
    \caption{Scatter plot between PM sensor output and reference instrument observations of the mobile monitoring data}
    \label{fig:scatterplot}
\end{figure}

The calibration with MLR and RF by utilising only a single variable $\mathcal{S}$ shows limited performance. This poor performance infers a weak correlation between the $\mathcal{S}$ and $\mathcal{Y}$, and the scatter plot  in Figure \ref{fig:scatterplot} clearly shows this weak positive correlation. This is a common scenario for mobile LCS due to the unwanted noise induced in mobile monitoring and the effect of meteorological parameters \cite{Arman}. When other covariates such as $T$ and $Rh$ are included in the calibration process along with $S$, both MLR and RF show improvements, which again follow the established trend in the literature since these variables influence the performance of the LCS \cite{maag_predep}.
 
Between MLR and RF models, the RF work better than the MLR for mobile data also. However, the maximum $R^2$ of 0.64 and the minimum $RMSE$ of 0.52 are able to achieve with the RF model with $\phi$ which are not attaining the guidelines of USEPA. According to USEPA at least $R^2$ of 0.8 between the LCS values and reference stations observations is required in order to accept the LCS data for the indication purpose. 

The proposed \textit{EEATC} is an attempt in that aspect and it is able to attain the benchmarks. In both training and testing, it is able to maintain the $R^2$ more than 0.8 and outperform the MLR and RF in all aspects shown in table \ref{tab:single_phase}. 

\begin{figure}[hbt!]
     \centering
     \subfloat[\label{fig:mobile_eeatc_train}]{%
        \includegraphics[width = \columnwidth]{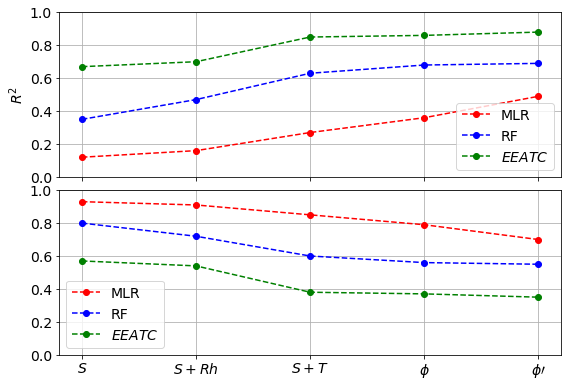}}
        \\
     \subfloat[\label{fig:mobile_eeatc_test}]{%
         \includegraphics[width = \columnwidth]{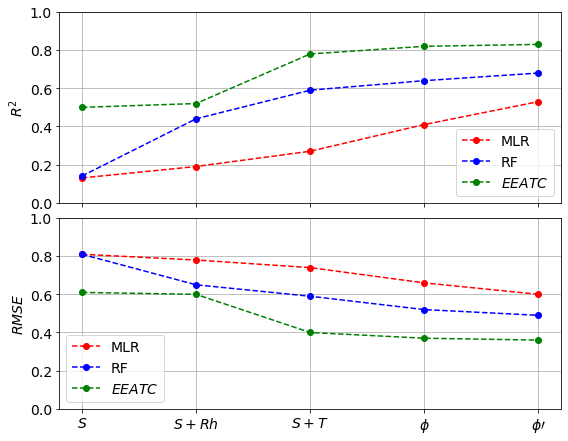}}
         \\
         \caption{comparison of \textit{EEATC} approach with MLR and RF in terms of $R^2$ and $RMSE$ for different features in the data sets, $\phi$ and $\phi \prime$ (a) in training (b) in testing }
\end{figure}

Further, we  propose an enhanced data set $\phi \prime$ for calibrating mobile LCS that encompasses the time-shifted input $\mathcal{S}(t-1)$ and the parameters included in the previously mentioned $\phi$ data set. The delayed versions of the inputs help to encounter the sensitivity effects and the temporal dependencies in the mobile measurements \cite{Esposito_dynamic_nn, HASENFRATZ2015268}. Our experimentation works with one unit delayed version, and the users need to test with different versions as per their setup. Therefore, $\phi \prime$ is the enhanced data set to calibrate the LCS in mobile deployments and it is represented in the matrix form as follows,
\begin{align*} \label{eq:enhanced_set}
\phi \prime = \begin{bmatrix}
\mathcal{S} & T & Rh & \mathcal{S}(t-1)\\
\end{bmatrix}_{(N \times 4)}
\end{align*}
The plots in Fig. \ref{fig:mobile_eeatc_train} and Fig. \ref{fig:mobile_eeatc_test} show the improvement in $R^2$ and $RMSE$ with $\phi \prime$ compared to $\phi$ in all the aspects.

\section{Conclusions and future work} \label{sec:conc_future}
The proposed \textit{EEATC} can calibrate the LCS in stationary and mobile deployments and outperform the well-known calibration models such as MLR and RF. This is proved empirically by comparing the performance of the \textit{EEATC} with the MLR and RF in terms of $R^2$ and $RMSE$ values. However, in the case of mobile deployments, it is tested on the data obtained through the pilot conducted in Chennai, India, for a limited period. Testing with different data sets obtained for various cities is a work for the future, and we believe that the \textit{EEATC} can work for this too. At this point, we would like to add a cautionary note that models and variables we presented  here may need to be suitably altered for different deployment conditions. 

Regarding the practical implementation of \textit{EEATC}, MLR constitutes the first phase and RF the second phase. That means the MLR first captures the linear relationships in the data, and then RF captures the non-linear part. Now, the order in which the models are cascaded is also an interesting aspect of \textit{EEATC} to explore. For example, how the method works when the RF is used as the first phase and MLR in the second phase. It can also be tested with different linear and non-linear models other than MLR and RF in future work. Further, using the same model or the same type of model in both phases of \textit{EEATC} may not work since they would capture only the linear or non-linear behaviour in the data, and this needs to be tested empirically in future work.  
\section{Resources}
Use the following GitHub link to access the codes to implement \textit{EEATC} and other models discussed in the article \href{https://github.com/mannamn-1/EEATC/upload/main}{{https://github.com/mannamn-1/EEATC/upload/main}}
\section*{Acknowledgments}
We thank Mr Aswin Giri, a graduate student, and Mr Vijay Kumar, a project engineer of Air Quality Lab, IITM, for their help during mobile monitoring

\bibliographystyle{IEEEtran}
\bibliography{reference}
\begin{IEEEbiography}[{\includegraphics[width=1in,height=1.25in,clip,keepaspectratio]{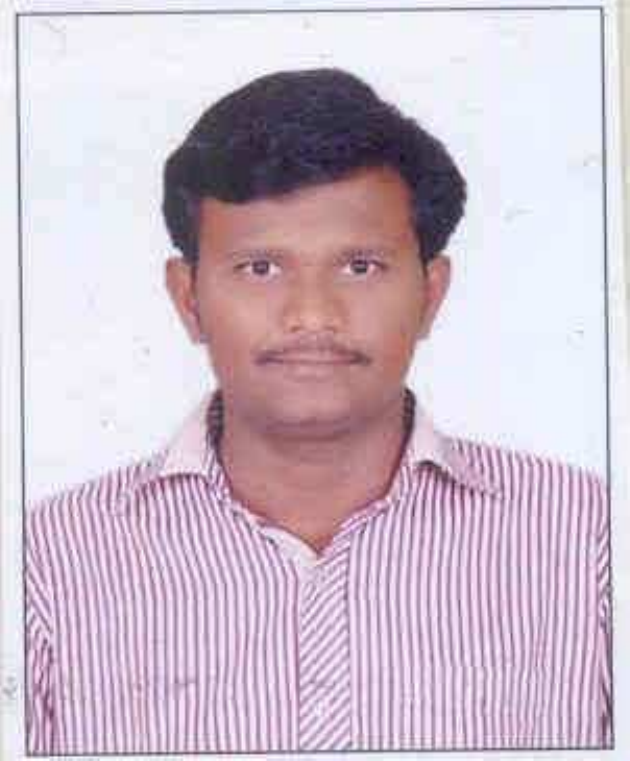}}]{M V Narayana} is currently a PhD. scholar in Electrical Engineering Department, Indian Institute of Technology Madras, India. He received his masters from Andhra University, Andhra Pradesh, India.  His research interests are sensor and Sensor Systems, Calibration models for sensors, IoT devices communication, wireless sensor networks, ML and AI. Currently, he is working on calibration models for low-cost sensors for air quality monitoring. 
\end{IEEEbiography}

\begin{IEEEbiography}[{\includegraphics[width=1in,height=1.25in,clip,keepaspectratio]{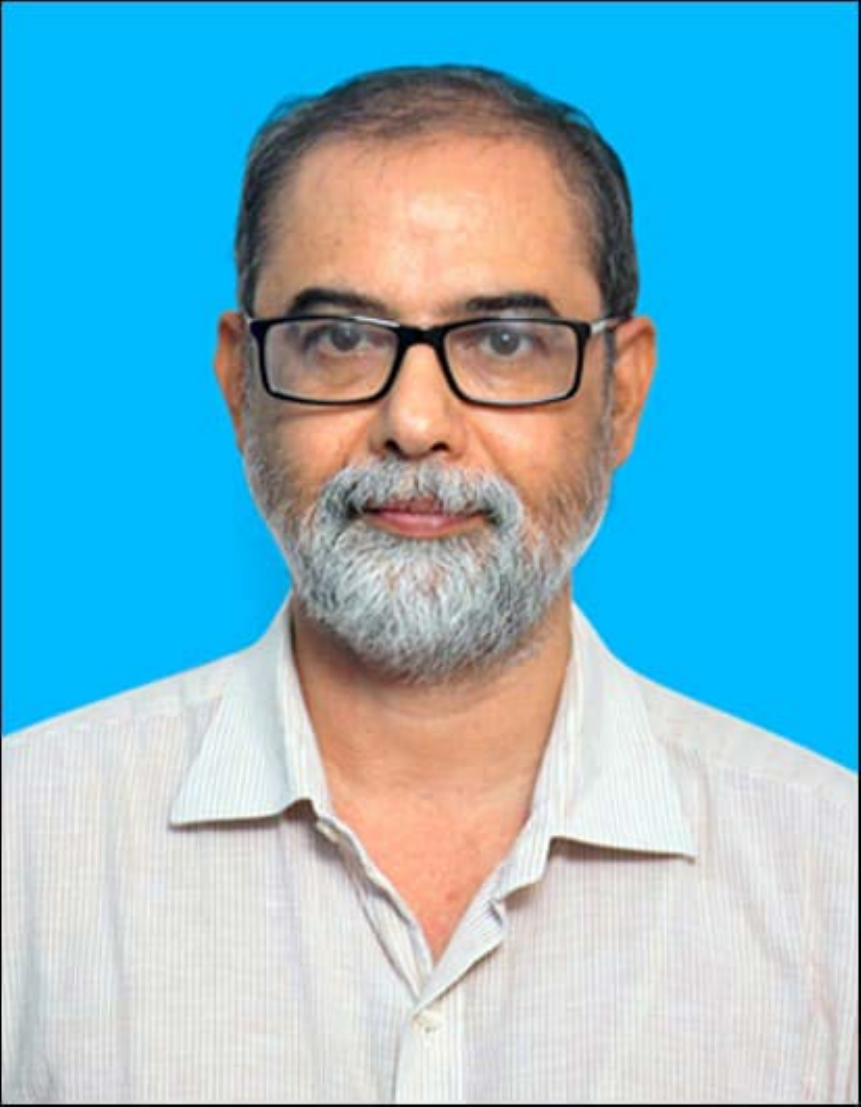}}]{Devendra Jalihal} is presently working as a professor in the Department of Electrical Engineering and Chairman of the Center for Continuing Education (CCE), Indian Institute of Technology Madras, Chennai, India. He served as head of the Electrical Engineering Department. He received PhD. from the University of Duke, Durham. He is a member of IEEE, IEICE and reviewer for IEEE, Elsevier, IET and IETE journals, and many conferences.  His research interests are statistical signal processing, detection and estimation theory, digital communication and wireless communication.
\end{IEEEbiography}

\begin{IEEEbiography}[{\includegraphics[width=1in,height=1.25in,clip,keepaspectratio]{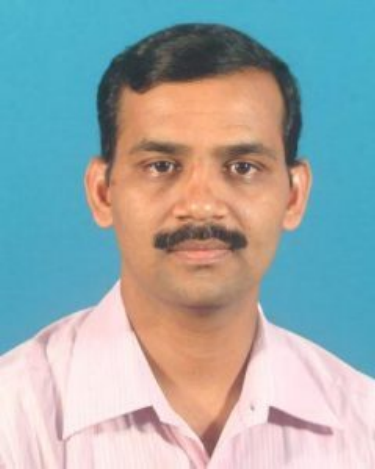}}]{Shiva Nagendra S M }is presently working as professor in Department of Civil Engineering, Indian Institute of Technology Madras, Chennai, India. He is author of books titled ‘Urban Air Quality Monitoring, Modelling and Human Exposure Assessment’ (ISBN:978-981-15-5511-4) and ‘Artificial Neural Networks in Vehicular Pollution Modelling’ (SCI-41, ISBN-10: 3-540-37417-5) published by Springer. He is associate editor of the journal Frontiers in Sustainable Cities, Frontiers and Journal of the Institute of Engineers (India): Series A, Springer. He is founder chairman of Indian International Conference on Air Quality Management (IICAQM) series and Vice-President, Society for Indoor Environment (SIE). His research interests focus on air quality management includes monitoring, source apportionment, modelling, design and development of emission control system, development of air quality management system and personal exposure monitoring.
\end{IEEEbiography}
\end{document}